\documentclass[%
reprint,
longbibliography,
amsmath,amssymb,
aps,
]{revtex4-2}
\usepackage{graphicx}
\usepackage{dcolumn}
\usepackage{bm}
\usepackage{amsmath}
\usepackage{amssymb}
\usepackage{algorithmic}
\usepackage{textcomp}
\usepackage{xcolor}
\usepackage{quantikz}
\usetikzlibrary{positioning}
\usepackage{adjustbox}
\usepackage{url}
\usepackage{booktabs}
\usepackage{float}

\begin{document}
	
\title{Improving Quantum Recurrent Neural Networks with Amplitude Encoding}

\author{Jack Morgan}
\affiliation{University of Chicago, Chicago, Illinois, USA}

\author{Eric Ghysels}
\affiliation{University of North Carolina, Chapel Hill, North Carolina, USA}

\author{Hamed Mohammadbagherpoor}
\affiliation{IBM T.\ J.\ Watson Research Center, Yorktown Heights, New York, USA}

\date{\today}

\begin{abstract}
	Recurrent Neural Networks (RNNs) are widely used to model complex temporal dependencies in sequential data. Quantum Recurrent Neural Networks (QRNNs) extend this idea to quantum machine learning by using quantum circuits inspired by classical recurrent architectures. In this paper, we propose three modifications to existing QRNN implementations that improve performance, generalization, and circuit efficiency. First, we introduce a simple preprocessing technique that augments amplitude encoded inputs with their prenormalized magnitudes, improving generalization on two financial data sets. Second, we show that a trained model can achieve comparable results with an approximate feature map rather than an expensive exact state preparation circuit. Third, we present a novel QRNN circuit architecture that is mathematically equivalent to the original model on ideal processors, but significantly reduces circuit depth. We demonstrate how these techniques can be put together to improve the performance of a QRNN when making predictions on real world financial data sets.
\end{abstract}
	
\maketitle
	
\section{Introduction \label{sec:intro}}
Recurrent Neural Networks (RNNs) are a class of machine learning models that are widely used for time series modeling and natural language processing tasks. There is a growing body of recent research on Quantum Recurrent Neural Networks (QRNNs) that replace the classical neuron structure with a parameterized quantum circuit (PQC).  \cite{PRXQuantum.4.020338} show that parameterized quantum circuits can be more expressive than classical models with a comparable number of parameters, which could lead to faster training convergence with smaller data sets for certain applications. The comparatively low energy consumption of quantum processors relative to the data centers used to train large neural networks may provide an additional incentive to study QRNNs, as discussed by \cite{10461108}.

Much of the existing QRNN literature focuses on hybrid approaches where the PQC serves as one subroutine in an otherwise classical model. For example, \cite{NEURIPS2020_0ec96be3} develop the Recurrent Quantum Neural Network (RQNN) with quantum neurons that use the amplitude amplification subroutine to create a nonlinear activation function. Similarly, \cite{9747369} introduce the Quantum Long Short Term Memory (QLSTM) architecture with five PQCs per time step.

As quantum hardware continues to develop, the study of near fault tolerant algorithms becomes increasingly relevant. One such model is the canonical QRNN proposed by \cite{LI2023148}, who built on \cite{PhysRevA.103.052414}. This model uses one PQC with a recurrent structure allowing for entirely quantum sequence modeling. The circuit contains a latent register which carries information between time steps, and a feature map register in which the data for each time step is encoded. Unlike \cite{PhysRevA.103.052414} who use a circuit Ansatz based on Hamiltonian dynamics, the canonical QRNN uses a hardware efficient unitary circuit that is more amenable to current devices. 

Executing QRNNs on quantum hardware introduces errors that can distort the loss landscape and interfere with parameter training. Canonical QRNNs suffer from coherence errors, which are the result of unobserved entangled qubits collapsing to one of the computational basis states. A similar circuit is used by other quantum models for time series such as Quantum Hidden Markov Models, Quantum Extreme Learning Machines, and Quantum Reservoir Computing techniques (e.g. \cite{ghysels2025quantum}, \cite{Xiong2025}, and \cite{chen2022reservoircomputingquantumrecurrent} respectively). \cite{ghysels2025quantum} provide a detailed analysis of how decoherence errors in the hidden register affect the performance of a quantum time series model. To mitigate this, \cite{LI2023148} propose a modification to the canonical QRNN circuit which they call Staggered QRNN or SQRNN. The principal idea of SQRNN is to shift the latent state register by one qubit each time step. Although this discards some information, it causes the latent qubits to be periodically reset which decreases the probability of decoherence errors.

A key component of the QRNN circuit is the feature map, which translates classical information into a quantum state. Most existing literature focuses on angle encoding, where each feature is mapped to a rotation angle on a dedicated qubit (see \cite{Chen_2024}, and \cite{Takaki_2021}). While efficient in time complexity, angle encoding requires $\mathcal{O}\left(N\right)$ qubits where $N$ is the number of features. 

In contrast, amplitude encoding requires only $\mathcal{O}(\log N)$ qubits. In the context of QRNN, reducing the number of feature map qubits leads to fewer parameters for two models with the same PQC structure and number of hidden states. Amplitude encoding has been shown to produce a more accurate model than angle encoding for a number of quantum machine learning tasks, including \cite{https://doi.org/10.1002/qute.202500611} and \cite{chen2025hybridquantumneuralnetworks}. Amplitude encoding also necessitates input normalization, which may distort sequence information. For example, two vectors at different magnitudes but the same direction would be encoded identically. This limitation, identified by \cite{chen2025hybridquantumneuralnetworks}, \cite{Li2025}, and \cite{West_2024}, remains an open problem in the literature. Despite its theoretical advantages, amplitude encoding is underexplored in fully quantum QRNNs. 

Amplitude encoding is less widely used than angle encoding for QRNNs because arbitrary quantum state preparation is computationally expensive. Amplitude encoding requires an arbitrary Quantum State Preparation (QSP) circuit. Exact QSP such as those from \cite{mottonen2004transformationquantumstatesusing} and \cite{PhysRevA.93.032318} require $\mathcal{O}(\exp(n))$ gates, where $n$ is the number of qubits. Alternatively, \cite{Nmaju2025}, \cite{10044235}, and \cite{Araujo2021} use ancilla qubits to reduce the circuit depth, however each method still scales exponentially in either the circuit depth or the requisite number of ancilla qubits. Other promising methods like \cite{PRXQuantum.5.030344} and \cite{10190145} provide speedups when preparing states with specific structure, however they are not suitable for the arbitrary state needed for encoding time series data. 

An alternative approach is to create an approximation of the desired amplitude encoded state. The primary categories of approximate state preparation are Matrix Product State (MPS) Encoding, Variational Encoding, and the Genetic Algorithm for State Preparation (GASP). MPS encoding techniques like \cite{BenDov2024} use a tensor network representation of the target state to decompose its preparation into a sequence of k-local operators. In general the size of said operators will be exponentially large. \cite{Creevey2023} introduce GASP which is based on the genetic algorithm classical optimization technique inspired by natural selection. Variational techniques train a PQC to prepare a state that is then fed into a subsequent model-specific PQC. As demonstrated by \cite{Nakaji_2022}, this approach is feasible but runs into some of the common quantum machine learning barriers such as barren plateaus. \cite{han2025enqodefastamplitudeembedding} introduce EnQode, a classical machine learning approach which prepares an approximate quantum state with a PQC Ansatz. Unlike traditional QSP, the time complexity of implementing EnQode is determined by a chosen Ansatz PQC. Common Ansatz circuits, including the one used by \cite{han2025enqodefastamplitudeembedding} and this paper, scale linearly with the number of qubits, i.e.\ with $\mathcal{O} (\log(N))$. EnQode uses a symbolic representation of the circuit to train the state preparation parameters. The benefit of this approach compared to other machine learning QSP algorithms (see \cite{PhysRevA.109.052423}, \cite{PhysRevResearch.4.023136}, \cite{wang2023robuststateboostingfidelityquantum}) is that exact gradients can be calculated with back propagation, and the parameters can be trained without repeatedly executing the circuit on quantum hardware. Errors in the state preparation routine will affect the downstream model performance. 
\cite{West_2024} found that a simple quantum neural network was able to maintain its accuracy at classifying images from standard data sets even with state preparation accuracy targets as low as 60\%. The impact of state preparation errors on QRNN accuracy remains to be seen.

These developments suggest that QRNNs with amplitude encoding may soon be practical and desirable. In this paper, we investigate three techniques designed to improve generalizability and hardware feasibility of such models. Namely, we:
\begin{itemize}
	\item[(a)] Introduce a prenormalized amplitude feature to preserve magnitude information lost in standard amplitude encoding.
	\item[(b)] Compare EnQode against exact amplitude encoding in the QRNN context.
	\item[(c)] Propose an alternating feature map register circuit structure that reduces quantum circuit depth required to implement a QRNN model.
\end{itemize}
Unlike the staggered QRNN of \cite{LI2023148}, which alters the model to reduce decoherence, the alternating $F$ register leaves the model unchanged and only restructures the circuit that implements it. To our knowledge, this work also provides the first evaluation of how approximate state preparation affects a trained recurrent model, in which the state preparation error is incurred at every time step.
For clarity, we refer to QRNNs using angle encoding, exact amplitude encoding, and EnQode based approximate amplitude encoding as Angle QRNN, Amplitude QRNN, and EnQode QRNN, respectively. The remainder of the paper is organized as follows. In Sec.~\ref{sec:background} we review the canonical QRNN and common data encoding schemes. In Sec.~\ref{sec:novelties} we outline the new methods we are introducing to improve generalizability. In Sec.~\ref{sec:methods} we describe the data and training procedures. In Sec.~\ref{sec:results} we explore the accuracy of the trained models, and the circuit depths of their different quantum circuits. In Sec.~\ref{sec:discussion} we interpret our findings and hypothesize additional fields of quantum machine learning that could adopt our techniques.

\section{Quantum Recurrent Neural Networks \label{sec:background}}
Before introducing the proposed novel techniques, we outline the classical RNN and the canonical QRNN from \cite{LI2023148}. Recurrent Neural Networks use a latent state which is dependent on the input data from the previous time step to make a prediction about future values. We borrow the notation used by \cite{LI2023148}, where said latent state $\vec{h}$ is used to make a prediction following the formalism:
\begin{align}
	\label{eq:rnn}
	\vec{y}_{t+1} &= f_y\left(U \vec{h}_t\right) \\
	\vec{h}_t &= f_h\left(V\vec{x}_t + W\vec{h}_{t-1}\right),
\end{align}
where $\vec{x}_t$ is the input of time step $t$, $f_y$ and $f_h$ are activation functions, whereas $U,$ $V,$ and $W$ are neural network weights to be optimized through the training process. 

In the quantum counterpart, the classical vectors $\vec{x}_t$ and $\vec{h}_t$ are replaced with the quantum state in two quantum registers: the feature map register $F$ and the hidden state register $H$, containing $n_F$ and $n_H$ qubits respectively. Increasing $n_H$ augments the expressibility of the model, which in turn increases the risk of overfitting. As alluded to in Sec.~\ref{sec:intro}, $n_F$ is determined by the number of features in the data and choice of encoding scheme. 

Each time step $t$ begins with a feature map $FM_t$ applied to the $F$ register to prepare the state $\ket{x_t}$ which is a function of the data $\vec{x}_t$. The exact form of $\ket{x_t}$ depends on the encoding method, which will be discussed later in this section. Next, a PQC Ansatz is applied to the entire circuit. We construct $A$ in a hardware efficient way using a combination of single qubit and two qubit gates native to IBM superconducting hardware. 
\begin{figure*}
			\begin{tikzpicture}
		\node(circ1){		
			\begin{adjustbox}{height=0.15\textwidth}
				\begin{quantikz}[column sep={0.4cm}, row sep={0.75cm,between origins}]
					\lstick{$\ket{0}_H$} & \gate[6]{A} & \\
					\lstick{$\ket{0}_H$} & & \\
					\lstick{$\ket{0}_H$} & & \\
					\lstick{$\ket{0}_F$} & & \\
					\lstick{$\ket{0}_F$} & & \\
					\lstick{$\ket{0}_F$} & & \\
				\end{quantikz}
			\end{adjustbox}
		};
		\node[right=of circ1](circ2){		
			\begin{adjustbox}{height=0.15\textwidth}
				\begin{quantikz}[column sep={0.4cm}, row sep={0.75cm,between origins}] 
					\lstick{$\ket{0}_H$} & \gate{R_Y\left(\theta_0\right)} & \gate{R_Z\left(\theta_6\right)} & \ctrl{1} & & & & & \gate{R_Y\left(\theta_{12}\right)} & \gate{R_Z\left(\theta_{18}\right)} \\
					\lstick{$\ket{0}_H$} & \gate{R_Y\left(\theta_1\right)} & \gate{R_Z\left(\theta_7\right)} & \targ{} & \ctrl{1} & & & & \gate{R_Y\left(\theta_{13}\right)} & \gate{R_Z\left(\theta_{19}\right)} \\
					\lstick{$\ket{0}_H$} & \gate{R_Y\left(\theta_2\right)} & \gate{R_Z\left(\theta_8\right)} & & \targ{} & \ctrl{1} & & & \gate{R_Y\left(\theta_{14}\right)} & \gate{R_Z\left(\theta_{20}\right)} \\
					\lstick{$\ket{0}_F$} & \gate{R_Y\left(\theta_3\right)} & \gate{R_Z\left(\theta_9\right)} & & & \targ{} & \ctrl{1} & & \gate{R_Y\left(\theta_{15}\right)} & \gate{R_Z\left(\theta_{21}\right)} \\
					\lstick{$\ket{0}_F$} & \gate{R_Y\left(\theta_4\right)} & \gate{R_Z\left(\theta_{10}\right)} & & & & \targ{} & \ctrl{1} & \gate{R_Y\left(\theta_{16}\right)} & \gate{R_Z\left(\theta_{22}\right)} \\
					\lstick{$\ket{0}_F$} & \gate{R_Y\left(\theta_5\right)} & \gate{R_Z\left(\theta_{11}\right)} & & & & & \targ{} & \gate{R_Y\left(\theta_{17}\right)} & \gate{R_Z\left(\theta_{23}\right)} \\
				\end{quantikz}
			\end{adjustbox}
		};
		\node at ($(circ1.east)!0.5!(circ2.west)$) {$=$};
	\end{tikzpicture}
	\caption{A diagram of an example efficient\_su2 Ansatz for an IBM processor. The unitary is constructed with a combination of two qubit entangling gates and parameterized native single qubit gates.}
	\label{fig:ansatz}
\end{figure*}
For all but the final time step, the $F$ register is reset to the initial state. The $H$ register carries information from the previous samples which will impact the prediction the QRNN will make for the final time step. In the last time step, a measurement is made in the $F$ register and the recorded probability corresponds to the prediction of the model. Figure~\ref{fig:QRNN_circuit} shows a high level QRNN circuit for two time steps, $n_H = 3$, and $n_F$=3.
\begin{figure*}
	\centering
	\begin{tikzpicture}[scale=1.0]
		\node[scale=1.0] (circ1) {
			\begin{adjustbox}{max size={0.55\textwidth}{\textheight}}
				\begin{quantikz}[column sep={0.5cm}, row sep={0.75cm,between origins}]
					\lstick{$\ket{0}_H$} & & \gate[6]{A} & & & \gate[6]{A} &\\
					\lstick{$\ket{0}_H$} & & & & & &\\
					\lstick{$\ket{0}_H$} & & & & & &\\
					\lstick{$\ket{0}_F$} & \gate[3]{FM_1} & & \midstick{$\ket{0}$} & \gate[3]{FM_2} & & \meter{} \\
					\lstick{$\ket{0}_F$} & & & \midstick{$\ket{0}$} & & & \\
					\lstick{$\ket{0}_F$} & & & \midstick{$\ket{0}$} & & & \\
				\end{quantikz}
			\end{adjustbox}
		};
		
		\node[right=1.2cm of circ1] (RNN) {
			\begin{adjustbox}{max size={0.4\textwidth}{\textheight}}
				\begin{tikzpicture}[
					node distance=1.1cm,
					every node/.style={font=\sffamily},
					input/.style={circle, draw, minimum size=10mm},
					hidden/.style={rectangle, draw, minimum size=10mm},
					output/.style={circle, draw, minimum size=10mm}
					]
					
					\node[hidden] (h0) {$h_0$};
					\node[hidden, right=of h0] (h1) {$h_{1}$};
					\node[hidden, right=of h1] (h2) {$h_{2}$};
					
					\node[input, below=of h1] (x1) {$x_{1}$};
					\node[input, below=of h2] (x2) {$x_{2}$};
					
					\node[output, right=of h2] (y3) {$y_{3}$};
					
					\draw[->] (h0) -- (h1);
					\draw[->] (h1) -- (h2);
					
					\draw[->] (x1) -- (h1);
					\draw[->] (x2) -- (h2);
					
					\draw[->] (h2) -- (y3);
					
				\end{tikzpicture}
			\end{adjustbox}
		};
		
	\end{tikzpicture}
	\caption{High level diagram of a canonical QRNN (left) and a classical RNN that inspired it (right). The $FM$ gates encode the input data for time step $i$ while the $A$ gate is a parameterized quantum circuit.}
	\label{fig:QRNN_circuit}
\end{figure*}
Data encoding is the process used to represent a certain data sample as a quantum state that can then be used by the PQC to make a prediction. Let $\vec{x}_t$ be a vector that represents one sample at time $t$, and $x_{i,t}$ for $i \in \{0,\dots,N-1\}$ be each of its features. We scale $x_{i,t}$ using the PyTorch \textit{MinMax} scaler to create $\hat{x}_t$ with features:
\begin{equation*}
	\hat{x}_{i,t} = \frac{x_{i,t} - \text{min}\left(x_i\right)}{\text{max}\left(x_i\right) - \text{min}\left(x_i\right)},
\end{equation*}
where $\text{max}\left(x_i\right)$ and $\text{min}\left(x_i\right)$ are the maximum and minimum values of feature $i$ respectively. A feature map is a quantum circuit which takes qubits that are typically initialized into the $\ket{0}$ state and outputs a quantum state that is a function of $\hat{x}_t$. 

Angle encoding, summarized by \cite{schuld2021supervisedquantummachinelearning}, is the most common data encoding approach for QRNNs. An angle encoding feature map requires one qubit per feature and uses Pauli rotations of an angle that is proportional to $\hat{x}_{i,t}$. We use the standard Pauli feature map from Qiskit \cite{javadiabhari2024quantumcomputingqiskit} with one repetition and $Y$ rotation gates. The resulting state takes the form:
\begin{equation*}
	\ket{x_t} = \bigotimes_{i=0}^{N-1} \left( \cos\left( \frac{\hat{x}_{i,t}}{2} \right) \ket{0}+ \sin\left( \frac{\hat{x}_{i,t}}{2} \right) \ket{1} \right).
\end{equation*}
Amplitude encoding encodes the values of $\hat{x}_t$ into the amplitudes of a quantum state over $\log_2(N)$ qubits using Quantum State Preparation (QSP). The resulting state is:
\begin{equation*}
	\ket{x_t} = \frac{1}{\alpha} \sum_{i=0}^{N-1} \hat{x}_{i,t} \ket{i},
\end{equation*}
where $\alpha$ is the $\ell_2$ norm of $\hat{x}_t$ and $\ket{i}$ denotes the $i^\text{th}$ computational basis state. The notation assumes $N$ is a perfect power of 2. When this is not the case, the first $N$ states have the aforementioned amplitude and the remaining ones have zero amplitude. This approach is exponentially more qubit efficient than angle encoding and has shown promise for larger data sets \cite{chen2025hybridquantumneuralnetworks}. Additionally, models trained with amplitude encoding have been shown to outperform comparable models with angle encoding on a variety of tasks (e.g. \cite{Nakaji_2022}, \cite{PhysRevA.109.052423}, \cite{chen2025hybridquantumneuralnetworks}, \cite{https://doi.org/10.1002/qute.202500611}, and \cite{Li2025}). However, exact algorithms for quantum state preparation such as \cite{mottonen2004transformationquantumstatesusing} and \cite{PhysRevA.93.032318} require circuits with exponential depth.

The two encodings place different demands on the hardware. Angle encoding requires $n_F = N$ feature qubits and a single layer of rotations per time step. Exact amplitude encoding requires only $n_F = (\log_2 N)$ feature qubits, but the state preparation circuits of \cite{mottonen2004transformationquantumstatesusing} and \cite{PhysRevA.93.032318} use $\mathcal{O}(N)$ CNOT gates and have depth $\mathcal{O}(N)$. EnQode also uses $(\log_2 N)$ feature qubits, and its Ansatz with $L$ layers has depth $\mathcal{O}(L\log_2 N)$. In practice, increasing the number of features may require additional layers to adequately capture the full range of the data. Consequently, $L$ depends on $N$ in an unquantified manner, suggesting that the depth of a suitable EnQode Ansatz may scale asymptotically faster than $\mathcal{O}(\log_2 N)$. For a fixed number of qubits and repetitions, a QRNN with angle encoding requires $\mathcal{O}(N)$ feature qubits and $\mathcal{O}(N)$ trainable parameters, while amplitude encoding requires $\mathcal{O}(\log_2 N)$ feature qubits and $\mathcal{O}(\log_2 N)$ trainable parameters.

The choice between angle and amplitude encoding is therefore application dependent. This conclusion was confirmed by \cite{https://doi.org/10.1002/qute.202500611}, who conduct a comparison using the same quantum machine learning model with different encoding schemes across different data sets. Angle encoding requires shallow circuits but a number of qubits, and hence of Ansatz parameters, that grows linearly with $N$. Amplitude encoding requires only logarithmically many feature qubits but a deeper state preparation circuit. Amplitude encoding is therefore more attractive in a few regimes that may or may not reveal themselves in practice. In a circumstance where prediction accuracy and training costs are top priorities, amplitude encoding may be selected due to the accuracy advantages reported in the literature. Alternatively, the number of available logical qubits could become a greater limiting factor than circuit depth on a forthcoming quantum processor. If this development occurs, then once again the qubit scaling advantage of amplitude encoding could be chosen over angle encoding. Finally, advancements in approximate state preparation like EnQode could alleviate the current circuit depth demands, which would weaken the principal advantage of angle encoding. The contributions of this paper concern how to improve a QRNN once amplitude encoding has been chosen, rather than establishing that amplitude encoding is preferable in general.

Next, we discuss some details about EnQode. Namely, another burgeoning area of study is the use of approximate methods for QSP. In particular, we are interested in the method proposed by \cite{han2025enqodefastamplitudeembedding} and coined EnQode. The algorithm begins with a k-means clustering of the normalized data set. They then train a PQC Ansatz to create the state corresponding to each centroid. The Ansatz used in the original EnQode paper, and our EnQode QRNN implementations, appears in Fig.~\ref{fig:EnQode_Ansatz}. Training is conducted classically with an exact symbolic representation of the state, which allows for efficient and accurate gradient calculations. Both of these steps occur `offline', or before  EnQode is used within a larger application. When generating a quantum circuit with a given sample, the algorithm identifies the centroid that is nearest to the new sample, then uses the saved parameters and gradients at this centroid to quickly learn how to prepare the new state. This process only needs to happen once for each sample in the data set, as the learned parameters can be used in future training epochs. This enables EnQode to generate approximate amplitude encoded states $\ket{\tilde{x}_t}$ with high fidelity (typically $\geq 0.9$), while maintaining logarithmic scaling in qubit count and, for a fixed number of Ansatz layers, in circuit depth. The fidelity between the approximate state and the true amplitude encoded state is defined by
\begin{equation}
	\text{fidelity} = |\braket{\tilde{x}}{x}|^2
\end{equation}
which ranges from 0 to 1. It is yet to be explored how the fidelity of the prepared samples impacts the training properties of a QRNN. 

\begin{figure*}
	\centering
	\begin{tikzpicture}[scale=1.0]
		\node[scale=1.0, midway](circ1){
			\begin{adjustbox}{max size={\textwidth}{\textheight}}
				\begin{quantikz}[column sep={0.5cm}, row sep={0.75cm,between origins}]
					\lstick{$\ket{0}_F$} & \gate{Rz(-\pi/2)} & \gate{Rz(\theta_0)} & \ctrl{1} & & \ctrl{1} & \gate{Rz(\theta_3)} & \ctrl{1} & & \gate{Rx(-\pi/2)} & \gate{Ry(-\pi/2)} \\
					\lstick{$\ket{0}_F$} & \gate{Rz(-\pi/2)} & \gate{Rz(\theta_1)} & \gate{Y} & \ctrl{1} & \gate{Y} & \gate{Rz(\theta_4)} & \gate{Y} & \ctrl{1} & \gate{Rx(-\pi/2)} & \gate{Ry(-\pi/2)} \\
					\lstick{$\ket{0}_F$} & \gate{Rz(-\pi/2)} & \gate{Rz(\theta_2)} & & \gate{Y} & & \gate{Rz(\theta_5)} & & \gate{Y} & \gate{Rx(-\pi/2)} & \gate{Ry(-\pi/2)} 
				\end{quantikz}
			\end{adjustbox}
		};
	\end{tikzpicture}
	\caption{High level diagram of the state preparation circuit Ansatz used by EnQode for 8 features (3 feature map qubits).}
	\label{fig:EnQode_Ansatz}
\end{figure*}
The probability $p$ of the measurement result at the end of the QRNN is mapped to a prediction using the inverse MinMax scaler, namely:
\begin{equation}
	\label{eq:mapping}
	y_{i,t+1} = x^{min}_i + p_{t+1} (x^{max}_i - x^{min}_i)
\end{equation}
where $y_{i,t+1}$ is the predicted value of the $i^{th}$ feature in the next time step, $x^{max}_i$ and $x^{min}_i$ are the largest and smallest values of the feature $i$ in the entire data set. For all models, including the Amplitude and EnQode QRNNs, we follow \cite{LI2023148} and set $p_{t+1}$ equal to the probability of measuring $\ket{1}$ on the first qubit of $F$. The models are trained to predict the daily return of the S\&P 500, so $i$ in Eq.~(\ref{eq:mapping}) refers to this feature.
\section{Proposed QRNN alterations}\label{sec:novelties}
We introduce three innovations designed to improve the performance of a QRNN that uses amplitude encoding. The first novelty is a classical preprocessing step in which we add a feature to the data set which captures the prenormalized amplitude of the original data vector. The second is the use of an approximate amplitude encoding scheme such as EnQode to remain feasible as the number of features increases, while incurring minimal loss of prediction accuracy on account of the state preparation error. The third modification is a new circuit layout that alternates data encoding between two feature map registers, which reduces the circuit depth and decoherence errors. Together, these changes aim to make amplitude encoded QRNNs more expressive and more viable on near term hardware.

\subsection{Preprocessing}\label{sec:preprocessing}
A key constraint of amplitude encoding is that all quantum states must be normalized, i.e., $\braket{x_t}{x_t} = 1$. This requirement removes magnitude information that could be relevant for sequence prediction tasks. For instance, a feature vector $\vec{x}_t$ and a scaled version $c \cdot \vec{x}_t$ (for any $c > 0$) are mapped to the same quantum state, as normalization erases any global amplitude. This information can be important for predictions in sequence modeling. This limitation of amplitude encoding is characterized in more detail by \cite{Li2025}.
To maintain amplitude information in the prepared state $\ket{x_t}$, we propose adding a feature to the data set that contains the $\ell_2$ norm of the original feature vector prior to normalization. This new feature needs to be added after the other features have been MinMax scaled, but before the vector has been normalized. We then scale the new prenormalized amplitude feature, and finally normalize the $N+1$ dimensional vector to create the state $\ket{x_t}$ to be encoded. 

\begin{figure}
	\centering
	
	\begin{minipage}{0.46\textwidth}
		\centering
		\includegraphics[width=\linewidth]{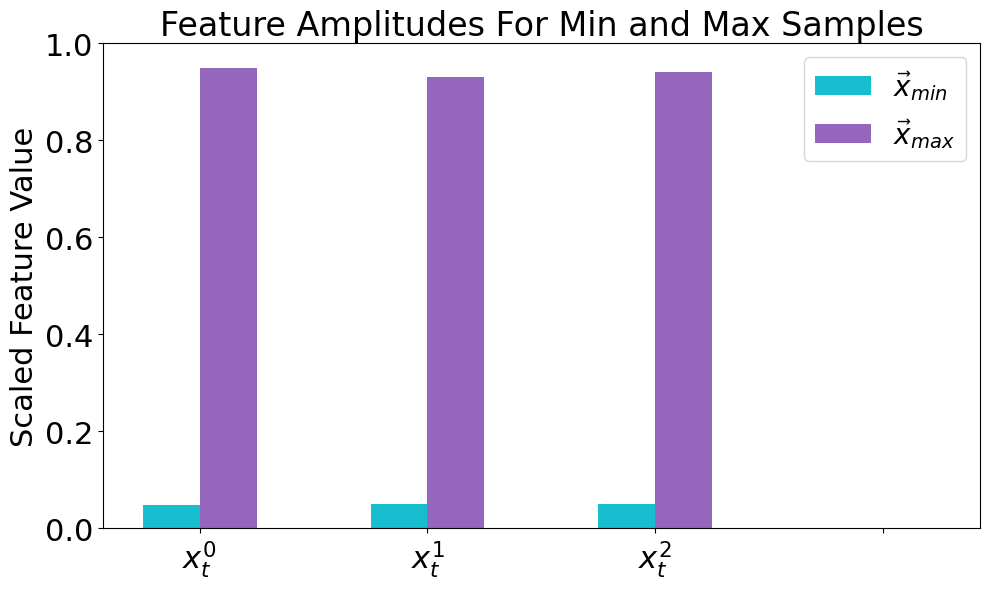}

	\end{minipage}
	\hfill
	\begin{minipage}{0.46\textwidth}
		\centering
		\includegraphics[width=\linewidth]{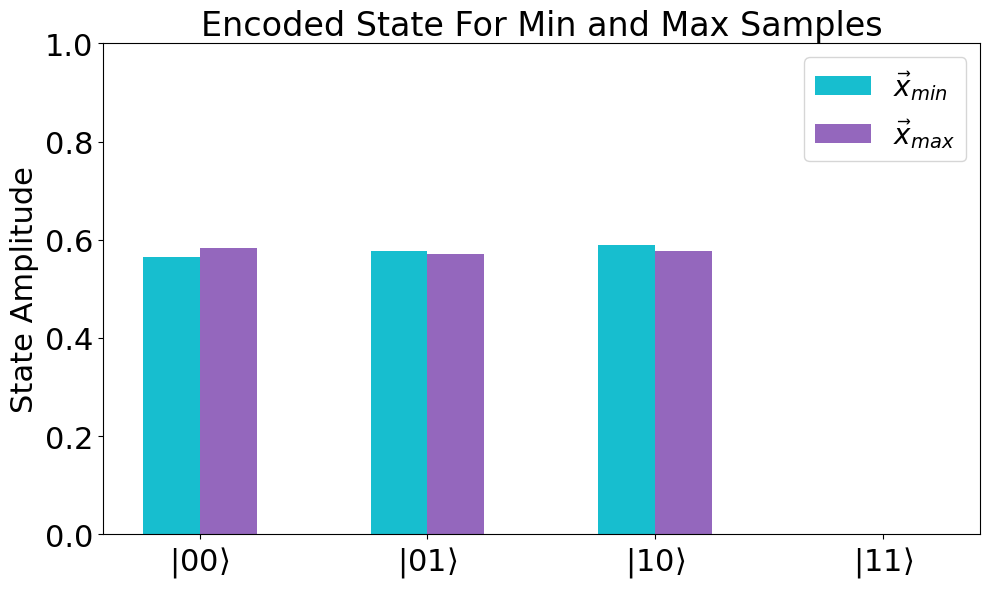}

	\end{minipage}
	
	\vspace{0.3cm}
	
	\begin{minipage}{0.46\textwidth}
		\centering
		\includegraphics[width=\linewidth]{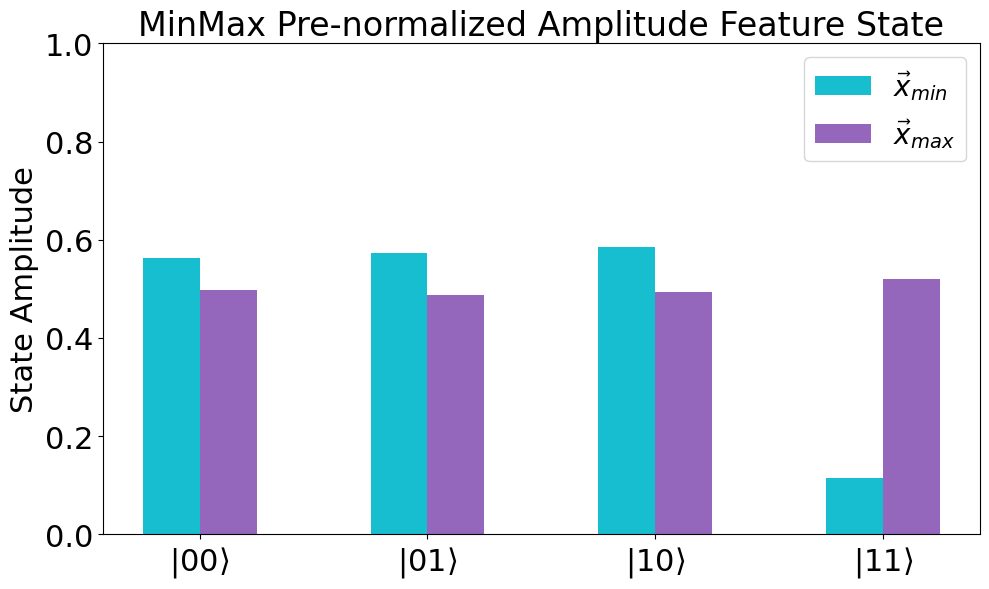}

	\end{minipage}
	\hfill
	\begin{minipage}{0.46\textwidth}
		\centering
		\includegraphics[width=\linewidth]{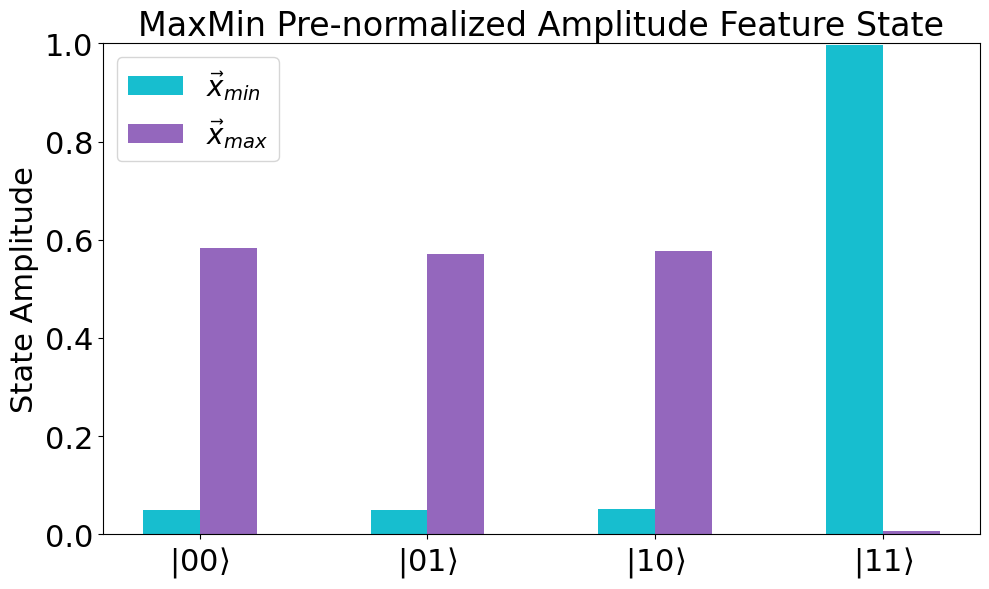}

	\end{minipage}
	
	\caption{Visualization of how each proposed preprocessing technique will translate the feature vectors for two different samples into quantum states. Without a prenormalized amplitude feature, the two vectors are indistinguishable to the circuit Ansatz. With a MinMax scaler, the sample with a larger feature $i$ value produces a smaller amplitude of the $\ket{i}$ state. With the MaxMin scaler, each day can be distinguished by the model, and individual feature relationships are maintained in their quantum representations.}
	\label{fig:pre-normalized_amplitude}
\end{figure}

For a system with $N$ original features, the augmented vector has $N+1$ dimensions. The added feature increases the overall norm, thereby reducing the relative amplitude of the original features. We provide an illustrative example of how the prenormalized amplitude feature affects the prepared quantum state in Fig.~\ref{fig:pre-normalized_amplitude}. Let $\vec{x}_{max}$ (high norm) and $\vec{x}_{min}$ (low norm) be two feature vectors that are in identical directions but differ in amplitude. The augmented vector ensures that $\vec{x}_{max}$ has a larger amplitude feature component, thus reducing the amplitudes of its other components compared to $\vec{x}_{min}.$ If we decide to scale only this added feature with a MaxMin scaler, then the $i^{th}$ component of $\vec{x}_{max}$ will be greater than that of $\vec{x}_{min}$. We hypothesize that using a MaxMin scaler for the amplitude feature will result in greater model generalizability.

\subsection{Feature Map}\label{sec:featuremap}
The recently introduced EnQode algorithm offers a scalable alternative to exact QSP for algorithms that require amplitude encoded inputs on large numbers of qubits. While it has been shown that the algorithm can prepare states with high fidelity, little research has been done into how the error in this state preparation will affect the resulting model generalizability. We explore this unknown area in the context of QRNNs by training the same model using both exact QSP algorithms and EnQode.

\subsection{Circuit Structure}\label{sec:circuit}
In the standard QRNN circuit, a single quantum register is reused across time steps. Between each step, the circuit must reinitialize and prepare a new input state while preserving the prior hidden state. This increases circuit depth, execution time, and the probability of decoherence errors.

We propose a new circuit structure with alternating $F$ registers. This allows the circuit to prepare $\ket{x_t}$ while the PQC processes $\ket{x_{t-1}}$. In Fig.~\ref{fig:AlternatingFRegisterQRNN} we show an example quantum circuit with the alternating $F$ registers. Unlike the staggered circuit innovation put forth by \cite{LI2023148}, a QRNN with an alternating circuit structure is mathematically identical to the canonical version on an ideal processor.

We can quantify the benefit of this restructuring in terms of $D_{FM}$, $D_A$, and $D_{\text{reset}}$, which denote the depths of the feature map, the Ansatz, and the reset operation respectively. Let $T$ be the sequence length, or number of times the encoding and Ansatz block is repeated. Ignoring routing overhead, the canonical circuit has depth $D_{\text{can}} \approx T(D_{FM} + D_A) + (T-1)D_{\text{reset}}$. In the alternating circuit, the reset and the next feature map run in parallel with the Ansatz, so its depth is $D_{\text{alt}} \approx D_{FM} + D_A + (T-1)\max\{D_A,\, D_{FM} + D_{\text{reset}}\}$. The depth saved is therefore
\begin{equation}\label{eq:depth_saving}
	D_{\text{can}} - D_{\text{alt}} \approx (T-1)\min\{D_A,\, D_{FM} + D_{\text{reset}}\},
\end{equation}
which grows linearly with the sequence length. For shallow feature maps such as EnQode, the saving is approximately $(T-1)(D_{FM} + D_{\text{reset}})$. The cost of the alternating register is $n_F$ additional qubits, which is only $\mathcal{O}(\log N)$ for amplitude encoding, together with additional SWAP gates on processors with limited connectivity.
\begin{figure}[H]
	\centering
	\begin{tikzpicture}[scale=0.75]
		\node[scale=1.0, midway](circ1){
			\begin{adjustbox}{max size={\textwidth}{\textheight}}
				\begin{quantikz}[column sep={0.5cm}, row sep={0.75cm,between origins}]
					\lstick{$\ket{0}_{F_b}$} & & \gate[3]{FM_2} & \gate[6]{A} & \midstick{$\ket{0}$} & & \\
					\lstick{$\ket{0}_{F_b}$} & & & & \midstick{$\ket{0}$} & & \\
					\lstick{$\ket{0}_{F_b}$} & & & & \midstick{$\ket{0}$} & & \\
					\lstick{$\ket{0}_H$} & & \gate[6]{A} & & & \gate[6]{A} &\\
					\lstick{$\ket{0}_H$} & & & & & &\\
					\lstick{$\ket{0}_H$} & & & & & &\\
					\lstick{$\ket{0}_{F_a}$} & \gate[3]{FM_1} & & \midstick{$\ket{0}$} & \gate[3]{FM_3} & & \meter{} \\
					\lstick{$\ket{0}_{F_a}$} & & & \midstick{$\ket{0}$} & & & \\
					\lstick{$\ket{0}_{F_a}$} & & & \midstick{$\ket{0}$} & & & \\
				\end{quantikz}
			\end{adjustbox}
		};
	\end{tikzpicture}
	\caption{High level diagram of the proposed QRNN circuit with alternating feature registers $F_a$ and $F_b$, shown for three time steps. While the Ansatz $A$ acts on $H$ and one feature register, the other feature register is reset and the next sample is encoded into it.}
	\label{fig:AlternatingFRegisterQRNN}
\end{figure}
\section{Training Methods}
\label{sec:methods}
To investigate how different preprocessing techniques, feature maps, and circuit structures affect the performance of canonical QRNNs on real world data, we train a variety of comparable models and evaluate the accuracy of their predictions. We repeat each training process with five different seeds and take the mean performance. We simulate all quantum models with 1024 shots on Qiskit's AerSimulator without hardware noise. For the feature map and circuit structure analysis, we use the noise model of the IBM Torino Heron r1 processor. All quantum models use Qiskit's efficient\_su2 Ansatz. The full source code used to generate all of our results is publicly available \cite{jackhmorgan2025QRNN}. 

We evaluate model performance using two data sets aimed at forecasting the daily return of the S\&P 500 (SPX). We chose 2017 because it is the most recent year for which the full Oxford-Man Institute realized volatility library is publicly available, courtesy of \cite{10.1162/REST_a_00300}. For the Oxford-Man data set, we use seven features. The first six are the daily return and median realized volatility for the S\&P 500, Dow Jones Industrial Average (DJIA), and Nasdaq indices. We also include the open to close returns for the S\&P 500 on the previous day. We also train with a data set sourced from Yahoo Finance. Our Yahoo Finance data set uses the daily return, daily high, and daily low values of the S\&P 500. The high and low values are represented as the log change from the previous day's closing price in order to maintain stationarity. We provide a complete list of features for both data sets in Table~\ref{tab:features}.
\begin{table}[t]
	\centering
	\begin{tabular}{c | c}
		Yahoo Finance & Oxford-Man \\
		\midrule
		S\&P Return & S\&P Return\\
		S\&P High & S\&P Realized Volatility\\
		S\&P Low & S\&P Open-Close\\
		 & DJIA Return\\
		 & DJIA Realized Volatility\\
		 & Nasdaq Return\\
		 & Nasdaq Realized Volatility\\
	\end{tabular}
	\caption{Table listing the features used for the Yahoo Finance and Oxford-Man data sets.}
	\label{tab:features}
\end{table}

In both cases we use sequences of 8 days and a test ratio of 0.2, which results in 192 training sequences and 51 testing sequences. With the prenormalized amplitude feature, the Yahoo Finance and Oxford-Man data sets contain four and eight amplitudes, which are encoded on two and three feature qubits, respectively. We train using the Mean Squared Error (MSE) of the measured probabilities relative to the measurement probability that maps to the correct value of $x_{t+1}$, defined by:
\begin{equation}
	MSE = \frac{1}{M} \sum_{m=1}^{M} \left(p^y_{t+1,m} - p^x_{t+1,m}\right)^2
\end{equation}
where $m$ indexes the $M$ sequences, $p^y_{t+1}$ is the measured probability that maps to the prediction in Eq.~\ref{eq:mapping}, and we can derive $p^x_{t+1}$ from the same equation and the known next value in the sequence $x_{i,t+1}$ using
\begin{equation}
	p^x_{t+1} = \frac{x_{i,t+1}-x^{min}_i}{x^{max}_i -x^{min}_i}.
\end{equation}
The MSE defined here and reported in Sec.~\ref{sec:results} is not the MSE of the predicted returns. To find the MSE of the predictions themselves, we would need to multiply the result by the square of the range of $x_i$.
We train with PyTorch's implementation of the Adam optimizer with a learning rate of 0.03. Originally proposed by \cite{kingma2017adammethodstochasticoptimization}, Adam is a gradient based optimization method that combines momentum and adaptive learning rates. We use the Simultaneous Perturbation Stochastic Approximation (SPSA) gradient algorithm introduced by \cite{880982} with a step size of 0.001. SPSA approximates the gradient by sampling a random direction in the parameter space, evaluating the loss of the model at two offset points. This requires exactly two circuit executions regardless of the number of parameters. The commonly used parameter shift rule is unattractive for QRNNs because the number of circuit executions required scales with the number of parameters and time steps, as shown by \cite{PhysRevA.103.052414}. We determined the step size and learning rate through a brute force search through reasonable candidates and chose the combination that resulted in the smoothest convergence behavior.

\section{Results}
\label{sec:results}
\begin{figure}
	\centering	
	\begin{minipage}[c]{0.48\textwidth}
		\centering
		\includegraphics[width=\linewidth]{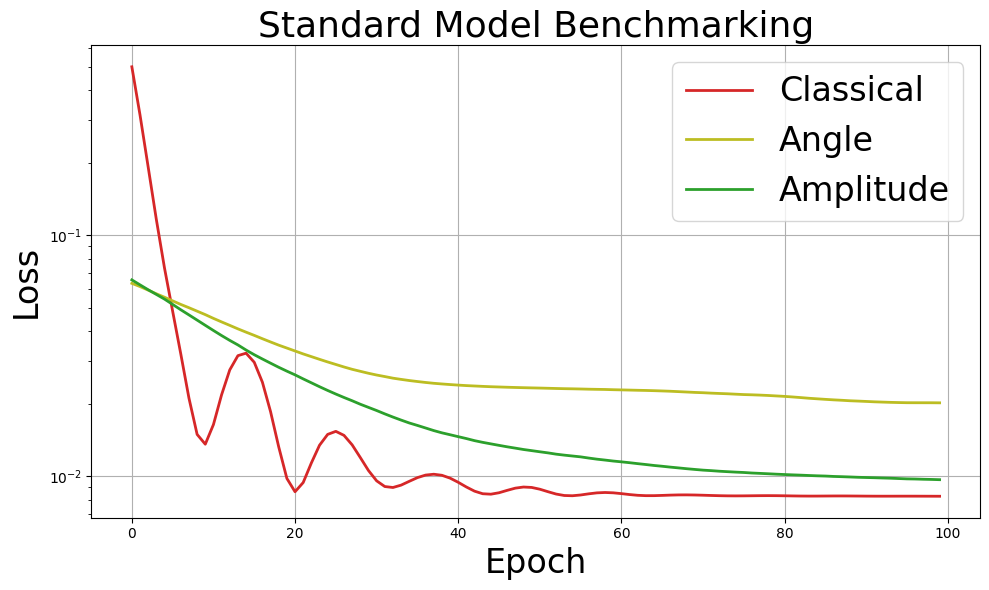}
	\end{minipage}
	\hfill
	\begin{minipage}[c]{0.48\textwidth}
		\centering
		\begin{tabular}{l c c}

			Model     & MSE    & Parameters \\
			\midrule
			Classical & 0.012  & 24         \\
			Angle     & 0.015  & 44         \\
			Amplitude & 0.009 & 28         \\

		\end{tabular}
	\end{minipage}
	\caption{Comparison between a classical RNN, an Angle QRNN, and an Amplitude QRNN without any of the proposed modifications. The training curves (left) and the test MSE and number of trainable parameters (right) are averaged over five training runs with different random seeds.}
	\label{fig:standard_models}
\end{figure}

\begin{figure}
	\centering
	
	\begin{minipage}[c]{0.48\textwidth}
		\centering
		\includegraphics[width=\linewidth]{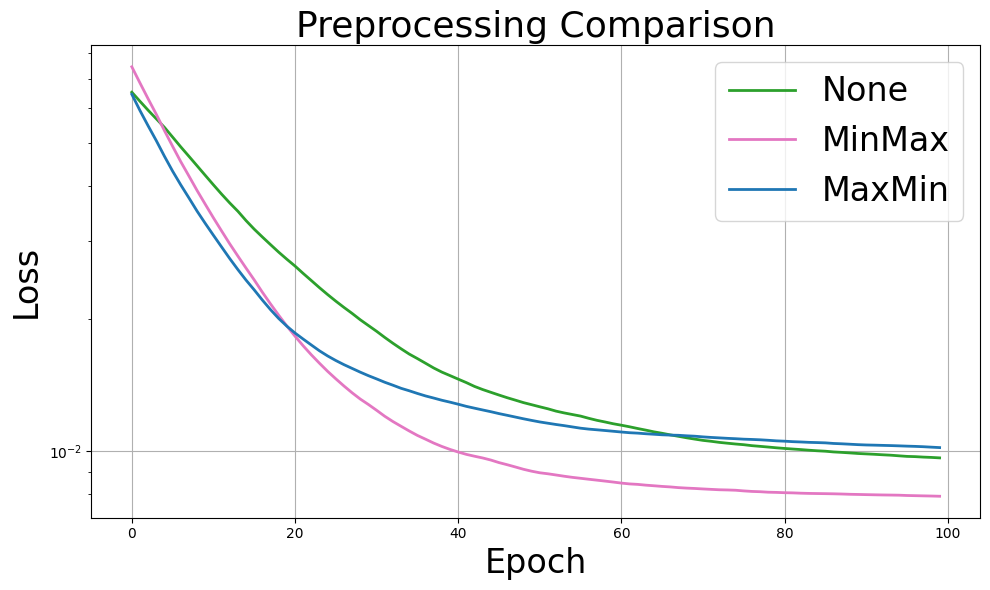}
	\end{minipage}
	\hfill
	\begin{minipage}[c]{0.48\textwidth}
		\centering
		\begin{tabular}{l c c}
			
			Preprocessing & MSE & MSE Ratio \\
			\midrule
			None & 0.0088 & 1.00 \\
			MinMax & 0.0067 & 0.76 \\
			MaxMin & 0.0056 & 0.64 \\ 
		\end{tabular}
	\end{minipage}
	\caption{Comparison of Amplitude QRNNs without a prenormalized amplitude feature (None), with a MinMax scaled amplitude feature (MinMax), and with a MaxMin scaled amplitude feature (MaxMin). The training curves (left) and the test MSE (right) are averaged over five training runs with different random seeds. The MSE ratio is the test MSE of each model divided by that of the model without the added feature.}
	\label{fig:preprocessing}
\end{figure}

We begin by benchmarking the base Amplitude QRNN against the Angle QRNN and an example classical RNN defined in Eq.~\ref{eq:rnn}. We chose an RNN with two hidden states to match the two hidden qubits in the quantum counterparts, and because the resulting model uses a comparable number of parameters to the Amplitude QRNN. Figure~\ref{fig:standard_models} shows the mean training curve across five trials for each model, the average mean squared error (MSE) on the test set, and the number of trainable parameters. Our results corroborate the literature by suggesting that the Amplitude QRNN achieves better generalization than the Angle QRNN when using the same number of hidden states. The chosen Ansatz parameter count scales with the qubits in registers $H$ and $F$ combined. The Amplitude QRNN contains fewer qubits in $F$, thus the resulting model uses fewer trainable parameters. For this specific task, we see the classical model achieving faster convergence and a lower training MSE. However, we verify that the Amplitude QRNN creates more accurate predictions of the testing data. 
Now that we have confirmed that the Amplitude QRNN offers potential benefits relative to the canonical model in \cite{LI2023148}, we turn our attention to the novel modifications presented in Sec.~\ref{sec:novelties}. The first proposed method we investigate is the addition of an amplitude feature during preprocessing. We compare an Amplitude QRNN with three different preprocessing procedures. The `None' model is the one seen in Fig.~\ref{fig:standard_models}, where we have done no extra preprocessing to the data. For the next two models we add a feature which corresponds to the prenormalized amplitude of the feature vector with each feature MinMax scaled. We test the model with this added feature `MinMax' scaled, and `MaxMin' scaled as denoted in Fig.~\ref{fig:preprocessing}.
While the MinMax scaled amplitude converged more quickly to the training data, we found that the MaxMin model was more generalizable with the test data set. 

\begin{figure}
	\centering
	
	\begin{minipage}[c]{0.48\textwidth}
		\centering
		\includegraphics[width=\linewidth]{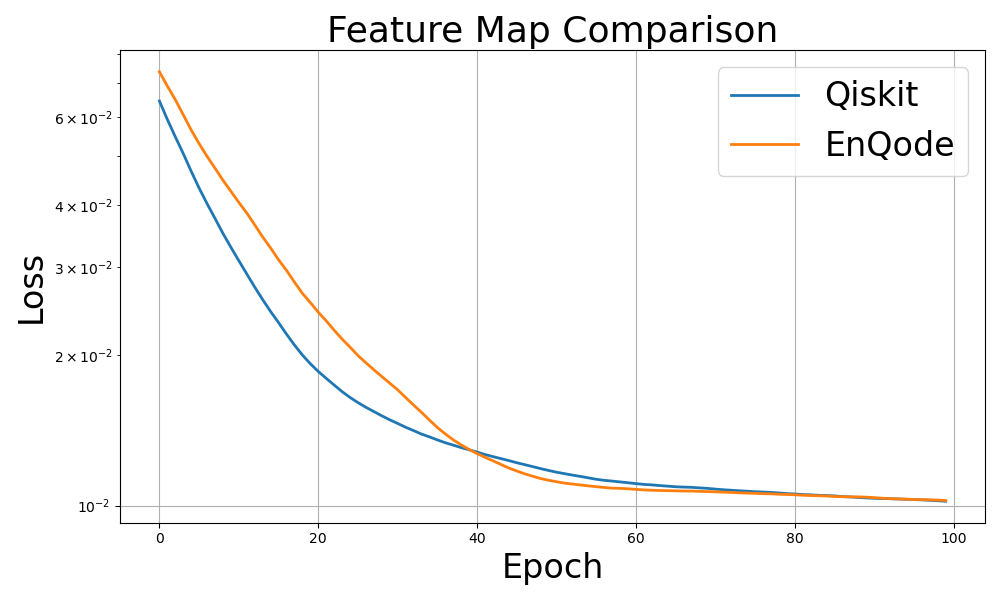}
	\end{minipage}
	\hfill
	\begin{minipage}[c]{0.48\textwidth}
		\centering
		\includegraphics[width=\linewidth]{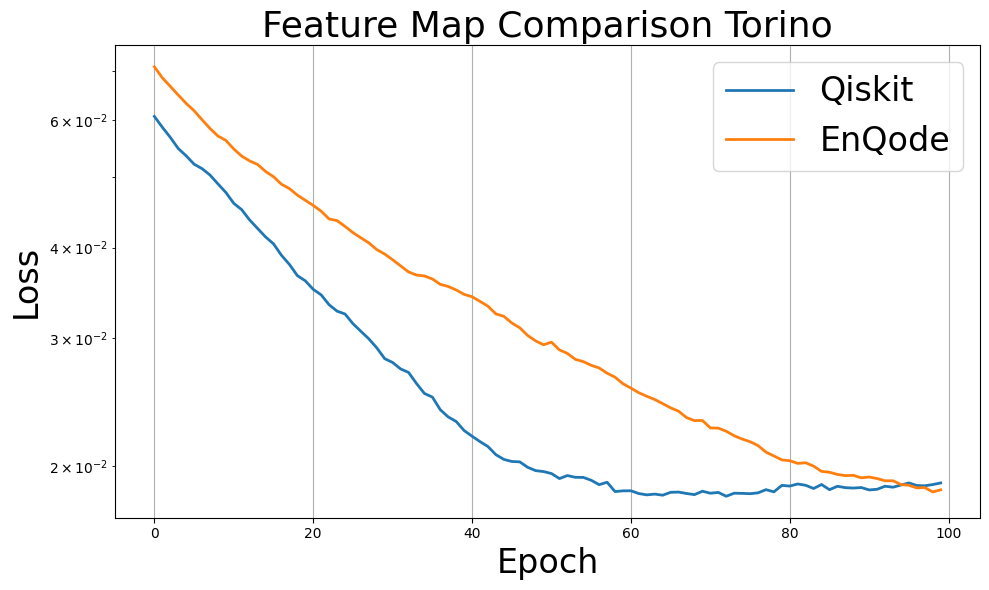} 
	\end{minipage}
	
	\vspace{1em} 
	
	\centering
	\begin{minipage}{0.5\textwidth}
		\centering
		\small
		\begin{tabular}{l | cc | cc}
			
			Noise Model & \multicolumn{2}{c}{None} & \multicolumn{2}{c}{IBM Torino}  \\
			\midrule
			Feature Map & MSE & MSE Ratio & MSE & MSE Ratio \\
			\midrule
			Qiskit & 0.0064  & 1.0 & 0.014 & 2.2 \\
			EnQode & 0.0090  & 1.4 & 0.013 & 2.0 \\
		\end{tabular}
	\end{minipage}
	
	\caption{Comparison between Amplitude QRNNs with the MaxMin scaled added feature using Qiskit's exact state preparation algorithm and EnQode. We present the average training curve without noise (left) and with the simulated noise of IBM Torino (right). In the table we present the test loss of each model as an exact value, and as a ratio relative to the test loss of the noiseless exact model.}
	\label{fig:data_encoding}
\end{figure}

In light of this result, we use the MaxMin scaled amplitude feature in the remaining experiments, beginning with a comparison of amplitude encoding techniques. Figure~\ref{fig:data_encoding} compares the same model using either Qiskit's exact state preparation algorithm \cite{PhysRevA.93.032318} or EnQode approximate state preparation. The latter option is feasible with a large number of qubits, however it prepares the state $\ket{x_t}$ with greater error. We use the EnQode Ansatz shown in Fig.~\ref{fig:EnQode_Ansatz} and outlined in detail in the original paper. The EnQode Ansatz recreated the ideal state to be prepared with an average fidelity of 0.94. This infidelity resulted in a 40\% increase in the testing loss on an ideal quantum simulator. When the noise of a real quantum processor is simulated, the two feature maps yield comparable test losses (0.013 for EnQode and 0.014 for exact state preparation). On current hardware, the cost of the approximation is offset by the reduced circuit depth and the associated decrease in errors. However, the MSE of both methods with simulated hardware noise is more than twice that of exact preparation without noise. 

\begin{figure}
	\centering
	\begin{minipage}[c]{0.48\textwidth}
		\centering
		\includegraphics[width=\linewidth]{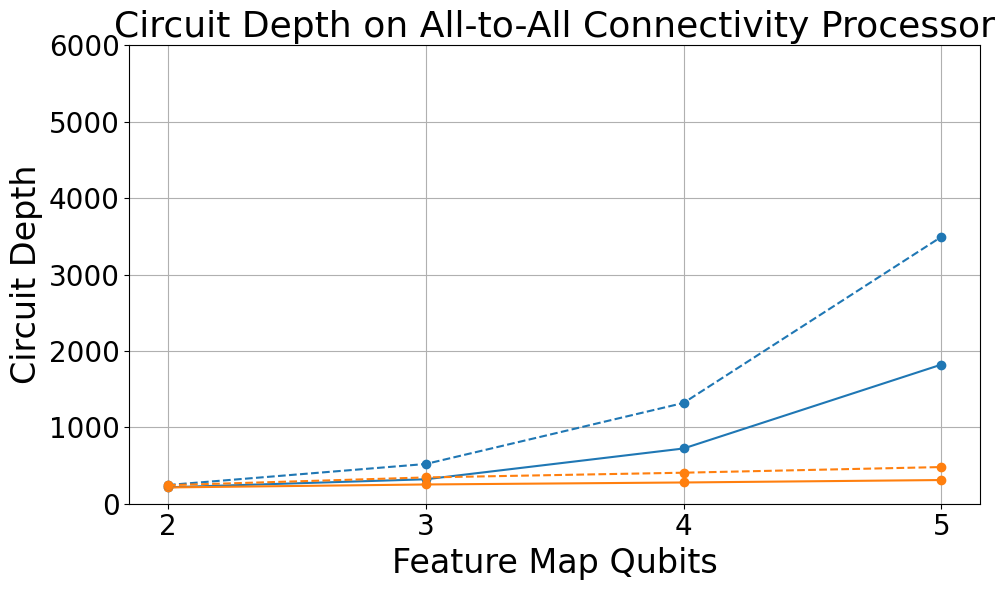}
	\end{minipage}
	\hfill
	\begin{minipage}[c]{0.48\textwidth}
		\centering
		\includegraphics[width=\linewidth]{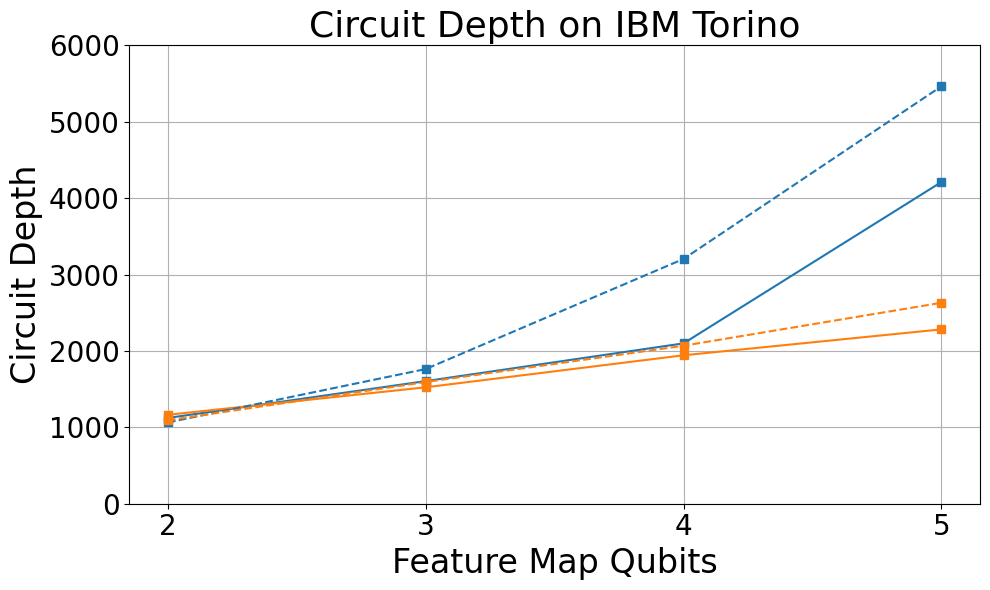}
	\end{minipage}
	\hfill
	\begin{minipage}[c]{0.48\textwidth}
		\centering
		\includegraphics[width=\linewidth]{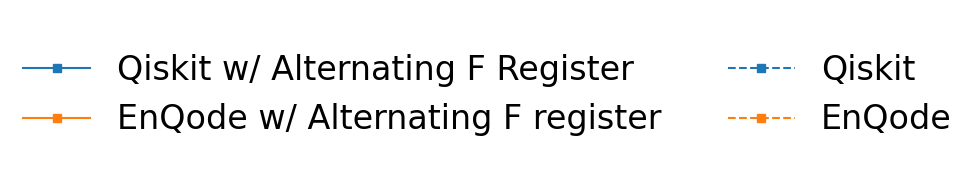}
	\end{minipage}
	\caption{The plot shows the circuit depth with Qiskit state preparation (blue) vs EnQode (orange), with and without (dashed and solid) an alternating $F$ register on an all-to-all connectivity processor and IBM Torino with an increasing number of $F$ qubits.}
	\label{fig:circuit_depths}
\end{figure}

We turn our attention to a comparison of the QRNN circuit with and without an alternating $F$ register. This innovation reduces the circuit depth and thus the execution time and the requisite coherence time of the $H$ register qubits. In our primary analysis, we use data sets with four and eight encoded amplitudes in order to quickly simulate the quantum circuits. At these problem sizes, the operations saved by the alternating $F$ register circuit are negligible compared to the overall circuit depth. In Fig.~\ref{fig:circuit_depths} we analyze how the depth of each circuit scales with additional features. As expected, the circuit depth with Qiskit's encoding algorithm grows exponentially while the EnQode circuit grows linearly. We see that the circuits constructed with the alternating $F$ register have lower depth than those without. As predicted by Eq.~\ref{eq:depth_saving}, the saving per time step is bounded by the depth of the feature map and reset, which grows with the number of feature qubits, until it reaches the depth of the Ansatz. The difference between standard architecture and that of our alternating $F$ register circuit increases with the number of feature map qubits. While the alternating $F$ register requires additional SWAP gates on limited connectivity processors, we still observe that the new circuit structure reduces the transpiled depth on IBM Torino.

The EnQode Ansatz in Fig.~\ref{fig:EnQode_Ansatz} uses two repetitions. In this exercise, we increase the Ansatz repetitions with the number of qubits. The authors of \cite{han2025enqodefastamplitudeembedding} provide a detailed algorithm for building the EnQode Ansatz with additional layers. Even with the additional Ansatz layers, the average simulated fidelity of the trained EnQode Ansatz decreases as we increase the dimension of the quantum state. In Tab.~\ref{tab:circuit_depths_enqode} we demonstrate how the circuit depths, fidelities, and number of features grow with the number of qubits.

\begin{table}
	
	\begin{minipage}[c]{0.48\textwidth}
		\centering
		\begin{tabular}{l c c c}
			Qubits & Features & Layers & Fidelity  \\
			\midrule
			2 & 4 & 2 & 0.94 \\
			3 & 8 & 3 & 0.91 \\
			4 & 16 & 4 & 0.83 \\
			5 & 32 & 5 & 0.74 \\
		\end{tabular}
	\end{minipage}
	\caption{This table shows the number of features that can be encoded, and the average classically computed fidelity between the ideal state $\ket{x_t}$ and the state prepared by the EnQode Ansatz (Fig.~\ref{fig:EnQode_Ansatz}) with the listed number of layers and qubits. The circuits measured in this table are used in Fig.~\ref{fig:circuit_depths} to calculate the depths of the complete QRNN circuit.}
	\label{tab:circuit_depths_enqode}
\end{table}

\section{Discussion}
\label{sec:discussion}
The authors of \cite{LI2023148} suggest that future work should explore how different encoding strategies impact the performance of their canonical QRNN. In this paper, we carried out such an analysis and proposed several innovations designed to improve model generalization and reduce circuit depth. We presented proof of concept simulations demonstrating that each proposed enhancement reduced either the test loss or circuit depth. First, we show that when using amplitude encoding, augmenting the input with a feature that corresponds to the prenormalized amplitude of each sample can help with model generalization. Second, we confirmed that a QRNN can use EnQode approximate amplitude encoding with a moderately reduced prediction accuracy and greatly reduced circuit depth, particularly as the number of features increases. Lastly, we introduced an alternating $F$ register quantum circuit that implements a mathematically equivalent QRNN with a shallower circuit structure.

Our analysis of QRNNs with amplitude encoding is motivated by a growing body of work comparing amplitude and angle encoding in quantum machine learning models, where the implementation with amplitude encoding has been shown to achieve a greater prediction accuracy. Whether these advantages persist across larger data sets remains an open question. While exact amplitude encoding is impractical at the scale of realistic data sets due to exponential state preparation costs, our results indicate that approximate preparation using EnQode can retain much of the predictive benefit while requiring circuits whose depth scales linearly with the number of qubits. Further analysis will be required at each scale to determine how the technique will scale to a realistic setting.

Increasing the sequence length increases the number of repeated QRNN blocks linearly, thereby increasing both the total execution time and the coherence requirements of the hidden state register. The computational overhead for both the proposed preprocessing method and EnQode feature map is unaffected by sequence length. The reduction in circuit depth provided by the alternating $F$ register applies to each time step equally. The total depth of each circuit, as well as the number of operations that are saved by adding the alternating $F$ register, scale linearly with the sequence length. According to Eq.~\ref{eq:depth_saving}, the depth saved by the alternating $F$ register for a sequence of length $T$ is approximately $(T-1)\min\{D_A, D_{FM} + D_{\text{reset}}\}$. The cost of training also scales favorably, since each SPSA gradient estimate requires two circuit evaluations regardless of the number of parameters or time steps. Deeper circuits and longer sequences may, however, make the model harder to train, since barren plateaus become more likely as the circuit grows~\cite{Nakaji_2022}.

More broadly, the techniques introduced in this paper are directly applicable beyond QRNNs to related quantum time series methods. The prenormalized amplitude feature could be used to augment any quantum model that encodes classical data with amplitude encoding, like \cite{Nakaji_2022}, \cite{PhysRevA.109.052423}, and \cite{chen2025hybridquantumneuralnetworks}. The alternating feature map layout can be applied to Quantum Reservoir Computing (QRC), Quantum Extreme Learning Machines (QELMs), and Quantum Hidden Markov Models (QHMMs). As discussed in \cite{chen2022reservoircomputingquantumrecurrent}, these approaches share a common circuit structure involving a latent register and repeated data encoding. Although prior work such as \cite{Xiong2025} and \cite{ghysels2025quantum} does not explicitly compare these architectures, their similarity suggests that the alternating circuit structure introduced here may transfer naturally to these settings. The impact of approximate encoding, prenormalized amplitude augmentation, and alternating register designs in these alternative frameworks remains an open area for investigation.

\bibliography{QRNN_BIB}

\end{document}